\documentclass[twoside]{article}
\usepackage{fleqn,espcrc2}

\usepackage{graphicx}
\usepackage[figuresright]{rotating}

\newcommand{\AmS}{{\protect\the\textfont2
  A\kern-.1667em\lower.5ex\hbox{M}\kern-.125emS}}

\title{Prospects for CP-violation searches in the 
       future experiment with RF-separated 
       $K^{\pm}$ beam @ U-70.}

\author{V.F.Obraztsov\address[IHEP]{Institute for High Energy Physics, \\        
        142284, Protvino, Moscow region, Russian Federation}%
        L.G.Landsberg\addressmark}

\begin{document}
 
\begin{abstract}
The first description  of the experimental program "OKA"
with RF-separated $K^{\pm}$ at Protvino U-70 PS is
presented. The parameters of the beam as well as the
potential for CP-violation searches are discussed.
\vspace{1pc}
\end{abstract}

\maketitle

\section{Introduction}
\begin{figure}[htb]
\vspace{9pt}
\includegraphics[width=6. cm]{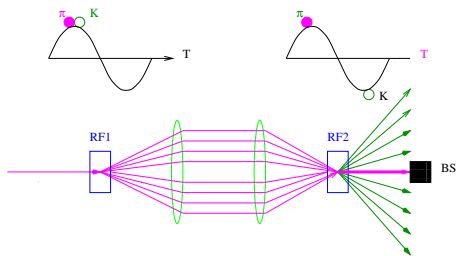} 
\caption{The principal scheme of RF separation.}
\label{fig:RF}
\end{figure}

If one looks at current (or recent) experiments with charged Kaons,
quite some with unseparated beams of different energies and intensities 
can be found, few with very low energy electrostaticly separated beams,   
but there is no high energy separated beams in operation. The reason is
that the relatively simple electrostatic $E \times B$ separation, which 
is very effective
at low energies does not work above few GeV and one has to build a high
power RF cavities which provide strong oscillating electrical field.
The most popular Panofsky-Montague-Schnell  scheme of separation is presented 
in Fig.\ref{fig:RF}.
The beam with definite momentum is sent to the system of two deflectors
with "-1" optics in between.
The phase shift between two deflectors is tuned in such a way that the
"unwanted" particles ($\pi$ in Fig.\ref{fig:RF}) have equal relative
phase in both deflectors. As a result, they get two deflections
which , in presented scheme, compensate each other and are
absorbed in the beam stopper. The distance between the deflectors is 
selected in such a way that Kaons, because of longer time-of-flight, have
$180^{\circ}$ shift with respect to pions, get double deflection and avoid
the absorber(it is interesting to note that because of random relation
$\frac{m_{p}^{2}-m_{\pi}^{2}}{m_{K}^{2}-m_{\pi}^{2}} \sim 4$, the protons are
suppressed "for free" together with pions) . \\
The RF deflectors must be superconductive to be able to keep the field
during slow extraction time of few seconds. Moreover, to get high 
quality factor of cavities it is necessary to cool them down to the
superfluid He temperature of 1.8 K$^{\circ}$. All these aspects make the
design technologically complicated. \\
At present, there are two projects which aim to build RF-separated high
intensity (slow extracted) beams:\\
1. FNAL Main Injector superconducting RF-separated beam for
 the CKM  experiment \cite{CKM}. \\
2. IHEP, Protvino  U-70 PS  superconducting RF-separated beam 
for the "OKA" experiment, based on existing Karlsruhe-CERN superconductive 
deflectors. 
\section{The Karlsruhe-CERN superconductive RF-deflectors.}
The only existing superconductive RF-separator was built at 
Karlsruhe(FRG) and used at CERN SPS at the end of $70^{es}$
\cite{RFNIM1}. It was an essential part of the S1 beam 
with  $K^{\pm}$, $\bar p$ of different energies for the 
$\Omega$-spectrometer.
The parameters of the deflectors are summarized in 
Table~\ref{table:karl}.
\begin{table}[htb]
\caption{The parameters of the CERN deflectors.}
\label{table:karl}
\begin{tabular}{|cc|}\hline
 Operating frequency(S-band) & 2865 MHz\\
 Wavelength, $\lambda$ & $\sim$ 10.5 cm\\
 Iris opening, 2a     & 40 mm\\
 Effective deflector length   & 2.74 m\\ 
 Number of cells/deflector    & 104\\ 
 Mean deflecting field        & $\sim$ 1.2 MV/m \\
 Working temperature          & T $\leq$ 2.5 K \\ \hline
\end{tabular}
\end{table}
In 1998, after the negotiations with CERN, the deflectors were transported
to Protvino. During 1998, the deflectors were connected to IHEP cryogenic
system and in summer, 1999 they passed complex RF tests at 4.2 K$^{\circ}$. 
It took one year more to build a cryogenic system capable to provide
 a temperature of T $\leq 2.5 K^{\circ}$.
 The final tests should start end of October 2000.
\section{The $K^{\pm}$ separated beam at U-70, based on
 Karlsruhe-CERN superconductive RF deflectors}
The goal of the "OKA" project ("OKA" is a Russian abbreviation 
for "Experiments with Kaons") is to build a high intensity RF-separated
$K^{\pm}$ beam based on the Karlsruhe-CERN superconductive deflectors using
the slow extracted 70 GeV proton beam of U-70 PS of IHEP,Protvino.
The project comprises: \\
1. Improvement of the slow extraction system of U-70. \\
2. Beam-line design and construction. \\
3. Design and construction of the cryogenic system for the 
superconductive RF-cavities with the working temperature of
 $1.8 \div 2.5 K^{\circ}$.\\
4. Physics program, Setup design and construction. \\
The schedule of the project assumes the start of the construction of the 
beam line in 2001. The first run with the kaon beam is scheduled for the 
spring of 2003. \\
The beam line design is completed, the parameters of the beam are presented in
the Table \ref{table:beam}.
\begin{table}[htb]
\caption{The parameters of the RF-separated beam.}
\label{table:beam}
\begin{tabular}{|lc|}\hline
 Target  & 50 cm Be\\
 Primary proton beam energy & 70 GeV\\
 Primary proton beam intensity & $10^{13}$ ppp\\
 Duty cycle                & 2/9 sec\\
 Spill/h                     & $4 \times 10^{2}$\\  
 Secondary beam momentum  & 12.5 or 18 GeV\\ 
 $\Delta$p/p $\%$ & $\pm 4$\\
 Horizontal acceptance   & $\pm 10$ mrad\\ 
 Vertical acceptance     & $\pm 1.9$ mrad \\
 Length of the beam line & $\sim$ 200 m \\ 
 Distance between separators & 76.3 m \\
 Intensity of $K^{+}(K^{-})$ at the end & $4(1.3)\times 10^{6}$ \\ 
 $\pi^{+},p$ contamination  & $< 50 \%$ \\ 
 Muon halo          & $ <100 \%$ \\ \hline
\end{tabular}
\end{table}
The~basic working energy is 12.5 GeV, it  is limited by the space available
in the U-70 gallery. The second working point of 18 GeV corresponds to the
condition of $360^{\circ}$ shift between $\pi$ and p and is supposed to be
used for the experiments with a target, for the hadron spectroscopy in the
Kaon beam, for example.\\
The intensity of the beam is limited by the relatively low deflection field
in the cavities ($\sim $1.2 MeV/m). As a consequence, the vertical acceptance
of the beam must be small. 
\begin{table*}[t]
\caption{The expected kaon decays statistics, 3 month, $\epsilon=50 \%$}
\label{table:decays}
\begin{tabular}{|c|c|c|c|c|}\hline
 Decay & Br&  acc. 
& { PDG}&
{  Expect.} 
\\ \hline
{   $K^{+}\rightarrow \mu^{+} \nu \pi^{0}$}&
{  $ 3.2\cdot 10^{-2}$} 
&{  0.27} &{  $ 10^{4}$} 
&{  $1.7 \cdot 10^{9}$} 
\\ \hline
{  $K^{+}\rightarrow  e ^{+} \nu_{e} \pi^{0}$}&
{  $ 4.8\cdot 10^{-2}$} 
&{  0.18} &{  $ 4 \cdot 10^{4}$} 
&{  $1.7 \cdot 10^{9}$} 
\\ \hline
{   $K^{+}\rightarrow \pi^{+} \pi^{+} \pi^{-}$}&
{  $ 5.6\cdot 10^{-2}$} 
&{  0.62} &{  $ 3 \cdot 10^{6}$} 
&{ $7. \cdot 10^{9}$} 
\\ \hline
{ $K^{+}\rightarrow \pi^{+} \pi^{0} \pi^{0}$}&
{  $ 1.7\cdot 10^{-2}$} 
&{ 0.17} &{  $ 10^{5}$} 
&{  $6 \cdot 10^{8}$} 
\\ \hline
{ $K^{+}\rightarrow  e^{+} \nu_{e} \pi^{+} \pi^{-} $}&
{  $ 3.9\cdot 10^{-5}$} 
&{  0.26} &{  $ 4 \cdot 10^{5}$} 
&{  $2. \cdot 10^{6}$} 
\\ \hline
{  $K^{+}\rightarrow  e^{+} \nu_{e} \pi^{0} \pi^{0} $}&
{  $ 2.1\cdot 10^{-5}$} 
&{  0.08} &{  $  35$} 
&{  $3 \cdot 10^{5}$} 
\\ \hline
{   $K^{+}\rightarrow \mu^{+} \nu_{\mu} \pi^{+} \pi^{-}$}&
{  $ 1.4\cdot 10^{-5}$} 
&{  0.62} &{  $ 7$} 
&{  $2 \cdot 10^{6}$} 
\\ \hline
{   $K^{+}\rightarrow e^{+} \nu $}&
{  $ 1.55\cdot 10^{-5}$} 
&{  0.45} &{  $ 10^{3}$} 
&{  $1.3 \cdot 10^{6}$} 
\\ \hline
{  $K^{+}\rightarrow \pi^{+} \gamma \gamma$}&
{  $ 1.1\cdot 10^{-6}$} 
&{  0.30} &{  31 } 
&{  $ 7 \cdot 10^{5}$} 
\\ \hline
{   $K^{+}\rightarrow \pi^{+}  \pi^{0} \gamma $}&
{  $ 2.8\cdot 10^{-4}$} 
&{ 0.21} &{ $ 2. \cdot 10^{4}$} &
{  $ 10^{7}$} 
\\ \hline
{   $K^{+} \rightarrow \mu^{+} \nu_{\mu} \gamma$}&
{  $ 5.5\cdot 10^{-3}$} 
&{ \ 0.4} &{ $ 2.5 \cdot 10^{3}$} 
&{  $4.3 \cdot 10^{8}$} 
\\ \hline
{  $K^{+}\rightarrow e^{+} \nu_{e} \gamma$}&
{  $ 3.8\cdot 10^{-5}$} 
&{  0.30} &{  $ \sim 100$} 
&{  $2.3 \cdot 10^{6}$ } 
\\ \hline
{  $K^{+}\rightarrow  e^{+} \pi^{0} \nu \gamma $}&
{  $ 2.6\cdot 10^{-4}$} 
&{  0.12} &{  $ \sim 250 $} 
&{  $6 \cdot 10^{6}$} 
\\ \hline
{   $K^{+}\rightarrow  \mu^{+} \pi^{0} \nu \gamma $}&
{  $ 2\cdot 10^{-5}$} 
&{  0.18} &{  $  $} 
&{  $7 \cdot 10^{5}$} 
\\ \hline
\end{tabular}
\end{table*}
\section{Experimental set-up} 

Four major U-70 setups: SPHINKS, GAMS, ISTRA, HYPERON plan to join the
"OKA" project. That gives a possibility to use existing equipment. \\
The upstream part of the setup is a beam spectrometer, with 
$\Delta P/P \sim 1 \%$, based on 1mm small gap PC's. In addition there are
3 beam $\check C$. The 12m long decay volume is surrounded by a photon
veto system ($\sim$ 1000 LG blocks $10 \times 10 \times 40 cm^{3}$).
The main magnetic spectrometer includes 3 Tm magnet
(200$\times$ 70 cm$^{2}$ aperture)
with 10K of PC's channels in front and 1K of 6cm $\phi$ mylar
straw tubes after. The set-up has two large aperture gamma-spectrometers:
GAMS-2000   downstream the straw's
and former EHS gamma detector  just after the decay volume,
used as large-angle detector. The central part of GAMS-2000 has 500 PWO
$2 \times 2 \times 20 cm^{3}$ crystals.
\section{A sketch of the experimental program.}
Let's make a brief estimate of the sensitivity of a "typical" experiment,
with the RF-separated beam:\\
With $\sim$ 13 $\%$ of 12.5 GeV Kaons decaying in the 12m decay volume we get
$5 \times 10^{5}$ {\small dec/spill}. In three month of data taking it gives 
$4\times 10^{11}$ decays. That is, the sensitivity of the experiment is    
$5 \times 10^{-11} \div 5 \times 10^{-12}$ for the detection efficiency 
of $\epsilon =0.05 \div 0.5$ .\\
Such a sensitivity is by an order of magnitude higher than that in proposed
experiments at the $DA\Phi$NE $\phi$-factory.\\
Table \ref{table:decays} shows the expected statistics for the main $K^{\pm}$
decays of interest. We have no possibility to discuss each line of the 
Table in details, let us just outline the main directions of the program:
\begin{enumerate}
\item Search for  direct CP violation in 
$K^{\pm} \rightarrow \pi^{\pm} \pi^{\pm} \pi^{\mp }( \tau)$ and 
$K^{\pm} \rightarrow \pi^{\pm} \pi^{0} \pi^{0 }( \tau')$ decays.
\item Search for  T-odd correlations in the decay 
$K^{\pm} \rightarrow \mu^{\pm} \nu \pi^{0} \gamma$.
\item Searches for S,P,T interactions in the decays 
$K^{+} \rightarrow e^{+} \nu$ , $K^{+} \rightarrow e^{+}(\mu^{+}) \nu \gamma$,
$K^{+} \rightarrow e^{+}(\mu^{+}) \nu \pi^{0} $.
\item  Low energy hadron physics in the decays 
 $K^{+} \rightarrow \pi \pi e^{+} \nu$, 
 $K^{+} \rightarrow \pi^{+} \pi^{0} \gamma$ etc.
\item Hadron spectroscopy and Primakoff physics with the $K^{\pm}$ beams.
\end{enumerate} 
In the following section we discuss in more details the points 1., 2. of
the program relevant for the topic of the present conference.
\section{Search for CP- violation in $\tau$ 
($K^{\pm} \rightarrow \pi^{\pm} \pi^{\pm} \pi^{\mp}$) 
and $\tau '$ ($K^{\pm} \rightarrow \pi^{\pm} \pi^{0} \pi^{0}$) decays.}
The direct CP violation would manifest itself by the 
difference between the $K^{+}$ and $K^{-}$ decay matrix elements if  
(see for example \cite{ModPhys}), there are at least two amplitudes with 
different "weak" phases  and different "strong" final state interaction
phases. The decays  $K^{\pm} \rightarrow \pi^{\pm} \pi^{\pm} \pi^{\mp}$
($\tau$) and $K^{\pm} \rightarrow \pi^{\pm} \pi^{0} \pi^{0}$ ($\tau '$)
satisfy these requirements. The direct CP could be detected, in particular,
by the observation of the difference $\frac{\delta g_{\pm}}{2g}$
in the slope parameters g$^{\pm}$ used in the conventional
parametrization for the $|M|^{2}$:
\begin{equation}
|M_{3\pi}|^{2} \sim 1+gY+hY^{2}+kX^{2}+.. 
\end{equation}
Here $Y= \frac{s_{3}-s_{0}}{m_{\pi^{2}}}$; 
$X=\frac{s_{2}-s_{1}}{m_{\pi^{2}}}$; 
$s_{i}=(p_{K}-p_{i})^{2}$, $s_{0}=\frac{1}{3}(s_{1}+s_{2}+s_{3});
\pi_{3}$ is the odd pion. The measured values of g are:
$g^{\tau} \sim .21$; $g^{\tau'} \sim 0.6$. \\
The theoretical predictions for the $\frac{\delta g_{\pm}}{2g}$ in the 
SM were originally spread in the wide range of $\frac{\delta g_{\pm}}{2g}= 
2  \times 10^{-6} \div 10^{-3}$ \cite{deltagSM}. Over last years they 
have converged to the value of $\sim 10^{-5}$ \cite{ModPhys},\cite{last}.
In the extensions of the SM, where CP violation is not constrained to the
CKM-matrix phases, larger values are possible. For example, in the three
Higgs doublet model of Weinberg \cite{3HDM}, where the origin of the CP
violation comes from the Higgs sector, the value of $\sim 2 \cdot 10^{-4}$
is predicted \cite{Shabalin}. 
\subsection{Existing experimental results}
There was only one old(1970!) dedicated experiment to measure the asymmetry
in the $\tau$ decay \cite{Ford}, based on the statistics of 
$\sim 1.5 \cdot 10^{6}$ events of each sign. The result is 
 $\frac{\delta g_{\pm}}{2g}= -0.70 \pm 0.53 \%$. \\
 The situation with the $\tau '$ is even worth: there was no dedicated
 experiment, the best measurement of $g^{-}$ is that of "ISTRA" setup
 at U-70, Protvino(40 K events): $g_{-}^{\tau'}=0,582 \pm 0,021$\cite{gISTRA}.
The best result for $K^{+}$ is from "HYPERON" setup at U-70(32K events):
$g_{+}^{\tau'}=0,736 \pm 0,02$\cite{gHYPERON}.
 Using these results, we get for the asymmetry:
$\frac{\delta g_{\pm}}{2g}= 0.1 \pm 0.02$- 5 $\sigma$ deviation from zero!

\subsection{Current experiments}
Before the discussion of the future results, it is useful to mention a 
simple formula which relates the statistical accuracy in the 
$\frac{\delta g_{\pm}}{2g}$ with the number of events:
\begin{equation}
\sigma(\frac{\delta g_{\pm}}{2g})= 
R \cdot \frac{\sqrt{n_{+}+n_{-}}}{2\sqrt{n_{+} \cdot n_{-}}};
\end{equation}
Here $n_{+}(n_{-})$ is the number of the decays in the $K^{+}(K^{-})$ beams;
R depends on g: $R_{\tau}=7.56$ ; $R_{\tau '}=3.0$. It is seen that 
$\tau '$ is factor of 6 more "effective".  
Now let us turn to the current experiments: \\
The first one is the HyperCP experiment at FNAL, which is aimed at searches
for CP violation in the hyperon decays \cite{HyperCP}. 
 The experiment had
accumulated 204M of $\tau^{+}$ and 75M of $\tau^{-}$ decays in the 1997 run,
which according to (2) corresponds to the statistical accuracy of 
$\frac{\delta g_{\pm}}{2g} \sim 6 \times 10^{-4}$. \\
The second experiment\cite{TNF} 
is the Tagged Neutrino Facility (TNF) at Protvino, U-70. The experiment is
using the tagging station of the TNF, which consists of  60 m decay volume,
3 large area hodoscope stations and an e.m calorimeter. The aim of the
first stage of the experiment is to accumulate $\sim$ 1M of $\tau '$ decay
of each sign. That corresponds to 
$\frac{\delta g_{\pm}}{2g} \sim 2 \times 10^{-3}$.   
\subsection{Future experiments}
There are two proposals, which aim to significantly improve the
 $\frac{\delta g_{\pm}}{2g}$ measurement. The first one is the NA48 proposal
 \cite{NA48}, and the second one is our proposal for the separated kaon
 beam at Protvino U-70. Table \ref{table:comp} shows the comparison of the
 two proposals. 
 \begin{table}[htb]
\caption{The comparison of the NA48 and 'OKA" proposals}
\label{table:comp}
\begin{tabular}{|ccc|}\hline
 Proposal & NA48 & OKA \\
 Beam & 5 $\% K^{\pm}$ & $>50 \% K^{+}$($K^{-}$)\\ 
 Beam p & 60 GeV & 12.5 GeV \\
 Duty cycle  & 5s/19.2s  & 2s/9s\\
 $K^{+(-)}$/cycle & $1.9(1.1) \cdot 10^{6} $& $4(1.3) \cdot 10^{6} $\\
 $\tau(\tau ') $eff. & 32 $\%$(50$\%$) & 50 $\%$(15$\%$) \\
 N $\tau^{+}(\tau^{-})$ & 1.3(0.75)$\cdot 10^{9}$ & 3.8(2.1) $\cdot 10^{9}$ \\  
 N $\tau'^{+}(\tau'^{-})$ & .82(0.47)$\cdot 10^{8}$ & 3.3(1.9) $\cdot 10^{8}$\\
 $\frac{\delta g}{2g} \tau (\tau ')$  & $1.7(3) \cdot 10^{-4}$ &
 $1.0(1.3) \cdot 10^{-4} $\\ \hline
\end{tabular}
\end{table} 
The advantage of the NA48 proposal is in the use of "simultaneous" $K^{\pm}$
beam, it allows to compensate the false asymmetry caused by the drift of the
set-up parameters in time, as well as to have the same trigger conditions
for the both beams. The advantage of the "OKA" is much lower total intensity
of the beam and larger acceptance. \\
The main difficulty of the proposed experiments is to demonstrate that,
indeed, there is no "dangerous" systematics at the level of $10^{-4}$.
In case of "OKA" a possible strategy of the measurements is the change of
the polarities of all the beam elements, say , every  6 hours. It's important
then to change the polarity of the main spectrometer magnet at the same time.
The possible sources of the systematics are: \\
1. Difference in the hadronic interactions between $\pi^{+}$ and $\pi^{-}$. \\
2. Different parameters of the positive and the negative beams. 
In particular, the tilt of the beams with respect to each other is 
potentially dangerous. \\
3. Effects of the resolution. \\
4. Variation of the setup and the beam parameters with time etc.... \\
We have started the detailed GEANT MC-studies of the different effects.
\begin{figure}
\vspace{9pt}
\includegraphics[width=6. cm]{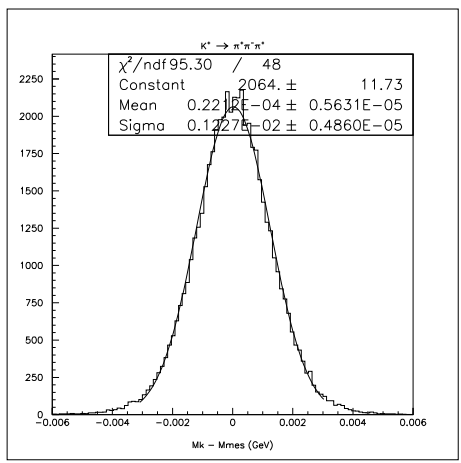} 
\caption{The  reconstructed $3\pi$ mass.}
\label{fig:burt1}
\end{figure}
\begin{figure}
\vspace{9pt}
\includegraphics[width=6. cm]{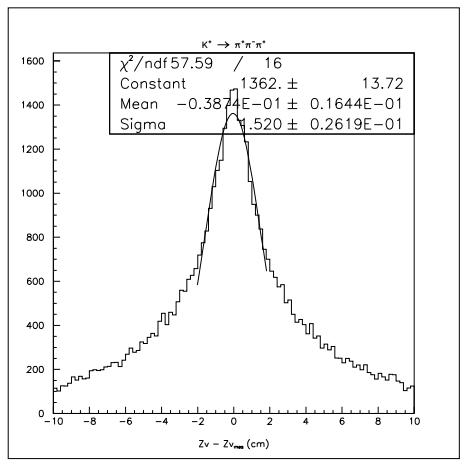} 
\caption{The  reconstructed z of the decay vertex.}
\label{fig:burt3}
\end{figure}
\begin{figure}
\vspace{9pt}
\includegraphics[width=6. cm]{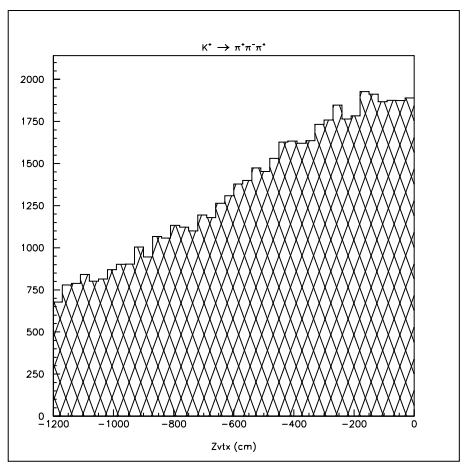} 
\caption{The acceptance of the setup versus z.}
\label{fig:burt4}
\end{figure}
\begin{figure}
\vspace{9pt}
\includegraphics[width=6. cm]{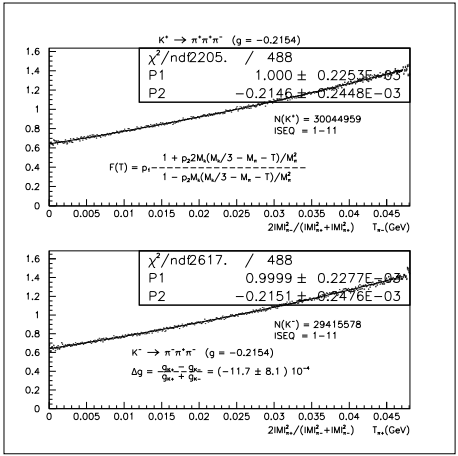} 
\caption{The events distribution over the kinetic energy
of the "odd" $\pi$ in the K-rest frame, normalized over
the same distribution for the "even" pion. The top figure is for the 
$K^{+}$, the bottom for the $K^{-}$. The difference in the slopes
$\frac{\delta g_{\pm}}{2g} < 10^{-3}$.}
\label{fig:burt5}
\end{figure}
The Fig.~\ref{fig:burt1} shows the reconstructed $3 \pi$ mass distribution,
the resolution is $\sim$~1.2 MeV; Fig.~\ref{fig:burt3}- the reconstructed
z-coordinate of the decay vertex(z -is directed along the beam). 
Fig.~\ref{fig:burt4} shows the acceptance of the setup versus z of the decay
point. It is seen that the decay volume length of $\sim$ 12m is kind of 
optimum. At last,  Fig.~\ref{fig:burt5} shows the events distribution 
over the kinetic energy of the "odd" pion in the kaon rest frame for the case
of $K^{+}$ and $K^{-}$ decays. The both distributions are normalized to that
for the "even" pions. These is the simplest way to remove the phase space
dependance. To extract the g-slopes, the distributions are fitted with the
formulae:
\begin{equation}
F(T) \sim \frac{1+2gM_{K}(M_{K}/3-M_{\pi}-T)/M_{\pi}^{2}}
{1+gM_{K}(M_{K}/3-M_{\pi}-T)/M_{\pi}^{2}}
\end{equation}
The denominator takes into account the "fake" slope for the "even" pion
distribution  which is $g/2$.
The last figure corresponds to the statistics of about 30M decays. 
It is seen that the difference in the slopes due to the
difference in hadronic interactions between $\pi^{+}$ and $\pi^{-}$
is smaller than $10^{-3}$. We'll continue these preliminary systematics 
studies in future, of course. What is interesting to mention, is that there
are several decays, where the $K^{\pm}$ asymmetry must be negligible. 
For example, in the decay $K \rightarrow e(\mu) \nu \pi^{0}$ the asymmetry
must be very small because there is practically no final state interaction.
The intense decay $K \rightarrow \pi \pi^{0}$ is another example, here the
expected effect is very small ($\frac{\Delta \Gamma}{2\Gamma}<10^{-9}$),
because the decay goes through  one amplitude (S-wave, $\Delta T=3/2$).
 These decays can be used to control the systematics.
\section{Search for T-violation in $K^{\pm}$ decays.}
The T-invariance is one of the fundamental symmetries in physics. It is
still an open problem to be tested experimentally with high precision.
T-violation must exist, because if CPT-invariance, which is considered 
by majority of theories as a "must" works, it is nothing more than
CP-violation which is already well established.  
T-invariance has been studied in many experiments. Typical example is
searches for the electric dipole moment of elementary particles
\cite {Dn}. Another possibility is to look for the T-odd correlations
in particles decays, such as triple correlations including spin in the neutron
decay.
A well known example of the T-odd correlation in the kaon decays is 
the polarization of the muon normal to the decay plane  
in $K^{+} \rightarrow \mu^{+} \nu \pi^{0}$ ($K_{\mu 3}$) decay:
\begin{equation}
\sigma^{\mu}_{\perp}= 
\frac{\vec \sigma_{\mu} \cdot (\vec p_{\pi} 
\times \vec p_{\mu})}{|\vec p_{\pi}\times \vec p_{\mu}|}
\end{equation} 
It was first suggested by J.J.Sakurai to look for T-violation by 
$\sigma^{\mu}_{\perp}$ measurement.
The unique feature of these process is that the "fake" T-odd correlation
due to the final state interaction (FSI) is very small: 
$\sigma_{\perp}^{FSI} \sim 5 \cdot 10^{-6}$ \cite{Zhit}.
Although the SM predicts zero effect, in some models (like 3HDM of Weinberg
\cite{3HDM}) a value higher than $10^{-3}$ is predicted. 
The best measurement of the $\sigma_{\perp}$, performed about 20 years
ago at the  BNL-AGS \cite{BNL} gives: $\sigma_{\perp} = 0.0031 \pm 0.0053$. 
Current result from KEK E246 \cite{E246} is 
$\sigma_{\perp} <10^{-2} @ 90 \%$ C.L., they hope to reach 
$\sigma_{\perp} <2.5 \cdot 10^{-3}$ at the end of the experiment.
\subsection {T-odd correlations in  $K \rightarrow \mu \nu \pi^{0} \gamma$
($e \nu \pi^{0} \gamma$).}
\begin{figure}
\vspace{9pt}
\includegraphics[width=6. cm]{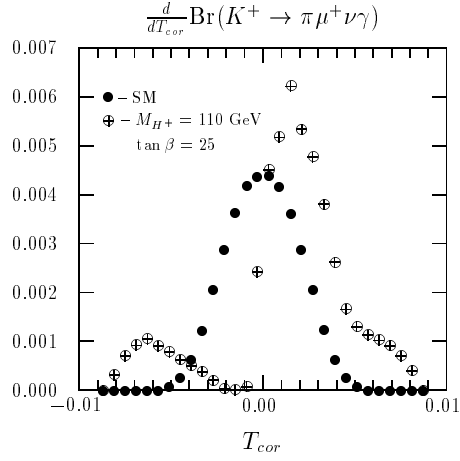} 
\caption{The  T-odd correlation for the 
$K \rightarrow \mu \nu \pi^{0} \gamma$ decay for the SM (black circles)
and 3HDM(open circles).}
\label{fig:ris1}
\end{figure}

We do not plan at present to measure $\sigma^{\mu}_{\perp}$ in the "OKA"
experiment: the energy of 12.5 GeV looks too high for that. On the contrary,
we are seriously considering the possibility of measuring the triple T-odd
correlation:
\begin{equation}
\label{eq:todd}
T_{\pi \mu \gamma}=\frac{\vec p_{\gamma} \cdot (\vec p_{\pi} 
\times \vec p_{\mu})}{|\vec p_{\gamma}|\cdot 
|\vec p_{\pi}\times \vec p_{\mu}|} 
\end{equation}
in the decay 
$K \rightarrow \mu \nu \pi^{0} \gamma$. 
Such a measurement was proposed
 for the first time by J.Gevais, J.Iliopoulous and J.Kaplan in 1966
\cite {Kaplan}. There are no theoretical predictions for the effect,
at present. The "fake" FSI correlation must not be strong: naively it must
be similar to $\sigma_{\perp}^{FSI}$ in $ K\rightarrow \mu \nu \gamma$,
which is  $\sim 10^{-4}$. The $K^{\pm}$ beam gives a principle possibility
of subtraction of the FSI: the FSI-correlation has the same sign in $K^{\pm}$,
while the true effect changes the sign. \\
The decay $ K\rightarrow \mu \nu \pi^{0} \gamma$ has not yet been observed,
the prediction of the SM is 
$Br(K\rightarrow \mu \nu \pi^{0} \gamma) \sim 2 \cdot 10^{-5}$. The only
experimental measurement of the T-odd correlation in the decay 
$K\rightarrow e \nu \pi^{0} \gamma$ is from ISTRA set-up at U-70 
\cite{T_ISTRA}
: $\frac{\vec p_{\gamma} \cdot (\vec p_{\pi} 
\times \vec p_{e})}{|\vec p_{\gamma}|\cdot 
|\vec p_{\pi}\times \vec p_{e}|}= 0.03 \pm 0.08 $; it is based on 192 
reconstructed events, which correspond to the 
Br=($2.7 \pm 0.2)\cdot 10^{-4}$ 
($E_{\gamma}^{c.m.}>10$MeV ; $0.6<cos\theta_{e,\gamma}<0.9$).\\
The IHEP theorists are performing detailed calculations of the 
correlation (\ref{eq:todd}) in SM and in the Weinberg's 3HDM
\cite{Likhoded}. This model is known to be a benchmark for the CP violation
searches beyond the SM. There are several parameters in the model
relevant for our case: $M_{H^{+}}$- the mass of the lightest charged Higgs;
$v_{1} ;v_{2} ;v_{3}$ - the VEV's of the 3 Higgs fields: 
$v_{1}^{2}+v_{2}^{2}+ v_{3}^{2}= 2 \sqrt{2} G_{F}$ ; 
$\alpha_{i}, \beta_{i}, \gamma_{i}$; i=1,2- 6 complex Yukawa coupling 
constants which are related to each other by the conditions: \cite{Cheng}
\begin{equation}
\frac{Im(\alpha_{2}\beta_{2}^{*})}{Im(\alpha_{1}\beta_{1}^{*})}=
\frac{Im(\beta_{2}\gamma_{2}^{*})}{Im(\beta_{1}\gamma_{1}^{*})}=
\frac{Im(\alpha_{2}\gamma_{2}^{*})}{Im(\alpha_{1}\gamma_{1}^{*})}= -1 
\end{equation}
\begin{equation}
  \frac{Im(\alpha_{1} \gamma_{1}^{*})}{v_{2}^{2}}= 
 -\frac{Im(\beta_{1} \gamma_{1}^{*})}{v_{1}^{2}} =
 -\frac{Im(\alpha_{1} \beta_{1}^{*})}{v_{3}^{2}}
\end{equation}
Further constraints on the model parameters can be extracted from the 
experimental data on $d_{n}$- the neutron electrical dipole moment; 
$\epsilon; \epsilon '$; $\Delta M_{D}$ etc.  
Given all the constraints, one still can expect a large value for  
$T_{\pi \mu \gamma}$. To illustrate, what one
can expect,  Fig.\ref{fig:ris1} shows the 
T-odd correlation, defined in a bit different way: 
$T_{cor}=\frac{\vec p_{\gamma} \cdot (\vec p_{\pi} 
\times \vec p_{\mu})}{m_{K}^{3}}$ for the reasonable set  of parameters 
in the 3HDM: $M_{H^{+}}=110$ GeV; $tan(v_{2}/v_{1})=25$; $v_{3}/v_{1}=10$;
$Im(\alpha_{1} \beta_{1}^{*})=0.8$, which is not excluded by any existing
measurement. A cut of $\theta{\mu \gamma}>20^{\circ}$ and $E_{\gamma}>30$ MeV
was done in the kaon rest frame to suppress the "trivial" IB part of the 
process. \\
In "OKA" we can detect  on the order of $7 \times 10^{5}$ events of the decay
$K^{+} \rightarrow \mu \nu \pi^{0} \gamma$, or $\sim 10^{5}$ events after the
kinematical cuts, which is quite sufficient to detect the effect of 
Fig.~\ref{fig:ris1}. Detailed simulation is in progress. 
\section{Conclusions}
The construction of the separated RF beam at U-70 PS of IHEP,Protvino will
allow to carry out a wide program of experiments, in particular, CP-violation 
searches in $K^{\pm}$ decays with a good sensitivity, competitive with 
other experiments. 

\end{document}